\documentclass[letterpaper,11pt]{article}
\pdfoutput=1
\usepackage{jheppub}
\usepackage{slashed}
\usepackage{graphicx}
\usepackage{subfigure}
\usepackage{soul}
\usepackage{amsmath}
\usepackage{multirow}
\usepackage{tablefootnote}
\usepackage{hyperref}
\usepackage{epstopdf}
\usepackage{lscape}
\usepackage{url}
\usepackage[dvipsnames]{xcolor}
\usepackage[utf8]{inputenc}

\newcommand{\beq}{\begin{equation}}
\newcommand{\eeq}{\end{equation}}
\newcommand{\bea}{\begin{eqnarray}}
\newcommand{\eea}{\end{eqnarray}}
\newcommand{\beqs}{\begin{subequations}}
\newcommand{\eeqs}{\end{subequations}}

\newcommand{\ba}{\begin{array}}
\newcommand{\ea}{\end{array}}

\def\figureautorefname~#1\null{Fig.\,#1\null}
\def\tableautorefname~#1\null{Tab.\,#1\null}

\def\equationautorefname~#1\null{Eq.\,(#1)\null}

\def\m1{M_1}
\def\m2{M_2}
\def\m3{M_3}

\def\ch10{\tilde \chi^0_1}

\def\gev{\,{\rm GeV}}

\def\to{\rightarrow}

\newcommand{\lsim}{\mathrel{\mathop{\kern 0pt \rlap
  {\raise.2ex\hbox{$<$}}}
  \lower.9ex\hbox{\kern-.190em $\sim$}}}
\newcommand{\gsim}{\mathrel{\mathop{\kern 0pt \rlap
  {\raise.2ex\hbox{$>$}}}
  \lower.9ex\hbox{\kern-.190em $\sim$}}}

\definecolor{pink}{RGB}{255,105,180}

\def\cosba{\cos(\beta-\alpha)}

\newcommand{\bpm}{\begin{pmatrix}}
\newcommand{\epm}{\end{pmatrix}}

\newcommand{\tanb}{\tan \beta}
\newcommand{\lambvs}{\sqrt{\lambda v^2}}

\allowdisplaybreaks[4]

\title{Probing loop effects in wrong-sign Yukawa region of 2HDM 
}

\author[a]{Wei Su}

\affiliation[a]{ARC Centre of Excellence for Particle Physics at the Terascale, Department of Physics,University of Adelaide, South Australia 5005, Australia}

\emailAdd{wei.su@adelaide.edu.au}

\abstract{ In the framework of 2HDM, we explore the wrong-sign Yukawa region with direct and indirect searches up to one-loop level. The direct searches include the latest $H/A \to f\bar f, VV, Vh, hh$ reports at current LHC, and the study of indirect Higgs precision measurements works with current LHC, future HL-LHC and CEPC. At tree level of Type-II 2HDM, for degenerate heavy Higgs mass $m_A=m_H=m_{H^\pm}<800$ GeV,
the wrong-sign Yukawa regions are excluded largely except for the tiny allowed region around $\cosba\in(0.2,0.3)$ under the combined constraints. The excluded region is also nearly independent of parameter $m_{12}$ or $\lambda v^2=m_A^2-m_{12}^2/(\sin \beta \cos \beta)$.
The situation changes a lot after including loop corrections to the indirect searches, for example $m_A=800 \gev$, the region with $\lambda v^2<0$ will be stronger constrained to be totally excluded. Whilst $\lambda v^2>0$ would get larger survived wrong-sign region, such as $\cosba \in (0,0.3)$ for $\lambvs=400 \gev$, under combined restrictions.}


\begin{document}
\maketitle
\flushbottom

\newpage
\section{Introduction and motivation}

Since the discovery of Standard Model (SM) -like Higgs boson at LHC Run-I~\cite{Aad:2012tfa,Chatrchyan:2012xdj}, SM is confirmed to be one self-consistent theory, and exploring Higgs boson properties especially Higgs couplings becomes a promising window to study new physics beyond-the-SM (BSM). Meanwhile motivated by various experimental and theoretical hits, to extend SM Higgs sector becomes necessary to address them.

Among numerous extensions, Two Higgs Doublet Model (2HDM) is a well motivated framework~\cite{Bernon:2015qea,Bernon:2015wef, Chun:2015xfx,Wang:2014sda}.  After electroweak symmetry breaking (EWSB), the general 2HDM will generate 5 mass eigenstates, a pair of charged Higgs $H^\pm$, one CP-odd Higgs boson $A$ and two CP-even Higgs bosons, $h, H$. Here we take the lighter $h$ as the measured SM-like Higgs. 

Since the improvements of various experiments, the wrong-sign region have attracted fruitful researches\cite{Ferreira:2014qda,Fontes:2014tga,Ferreira:2014dya,Biswas:2015zgk,Greiner:2015ixr,Modak:2016cdm,Ferreira:2017bnx,Han:2017etg,Coyle:2018ydo}. This work focuses on testing the so-called wrong-sign Yukawa region up to one-loop level with both indirect and direct searches at current LHC.  For the direct searches, we constrain the parameter space with various heavy Higgs decays, taking the cross section times branching ratio $\sigma \times Br$ limits of various channels, including 
$A/H \to \mu\mu$ \cite{CMS:2016tgd, Sirunyan:2019tkw, Aaboud:2019sgt}, 
$A/H \to bb$ \cite{Sirunyan:2018taj,Aad:2019zwb}, 
$A/H \to \tau\tau$ \cite{CMS:2017epy,Sirunyan:2018zut, Aaboud:2017sjh}, 
$A/H \to tt$ \cite{Sirunyan:2019wph,Aaboud:2019roo}, 
$H \to ZZ$ \cite{Sirunyan:2018qlb, Aaboud:2017rel}, 
$H\to WW$ \cite{CMS:2016jpd, Aaboud:2017gsl} at tree level. For the indirect searches, we perform the global fit the SM-like Higgs precision measurement from LHC Run-II \cite{ATLAS:2019slw}, HL-LHC \cite{Cepeda:2019klc} and CPEC \cite{CEPCStudyGroup:2018ghi} up to one-loop level. The results show that the wrong-sign Yukawa region for $m_A<800 \gev$ is strongly constrained. But the constraints get weaker afer including the loop correction to Higgs precision studies for $\lambda v^>0$, while for $m_A=1500 \gev$, the allowed $\cosba$ is smaller compared to it at tree level.

Our paper is structured as follows. In \autoref{sec:2hdm}, we will give a brief introduction to 2HDMs, concentrated on the wrong-sign Yukawa analysis. We give a brief summary of study methods and the relevant experimental reports in \autoref{sec:study_method}. Then at \autoref{sec:results_tree} and \autoref{sec:results_loop} we present our analyses and results at tree and one-loop level respectively. Finally we will give our main conclusions in \autoref{sec:con}.


\section{Two Higgs doublet models}
\label{sec:2hdm}
\subsection{2HDM Higgs sector}

The general 2HDM has two ${\rm SU}(2)_L$ scalar doublets $\Phi_i\ (i=1,2)$ with hyper-charge $Y=+1/2$,
\begin{equation}
\Phi_{i}=\begin{pmatrix}
  \phi_i^{+}    \\
  (v_i+\phi^{0}_i+iG_i)/\sqrt{2}
\end{pmatrix}\,.
\end{equation}
where $v_i\ (i=1,2)$ are the vacuum expectation values (vev) of the two doublets after EWSB with $v_1^2+v_2^2 = v^2 = (246\ {\rm GeV})^2$ and $\tanb=v_2/v_1$.

The 2HDM Lagrangian for the Higgs sector can be written as
\begin{equation}\label{equ:Lall}
\mathcal{L}=\sum_i |D_{\mu} \Phi_i|^2 - V(\Phi_1, \Phi_2) + \mathcal{L}_{\rm Yuk}\,,
\end{equation}
with a Higgs potential of
\begin{eqnarray}
\label{eq:L_2HDM}
 V(\Phi_1, \Phi_2) &=& m_{11}^2\Phi_1^\dag \Phi_1 + m_{22}^2\Phi_2^\dag \Phi_2 -m_{12}^2(\Phi_1^\dag \Phi_2+ h.c.) + \frac{\lambda_1}{2}(\Phi_1^\dag \Phi_1)^2 + \frac{\lambda_2}{2}(\Phi_2^\dag \Phi_2)^2  \notag \\
 & &+ \lambda_3(\Phi_1^\dag \Phi_1)(\Phi_2^\dag \Phi_2)+\lambda_4(\Phi_1^\dag \Phi_2)(\Phi_2^\dag \Phi_1)+\frac{\lambda_5}{2}   \Big[ (\Phi_1^\dag \Phi_2)^2 + h.c.\Big]\,,
\end{eqnarray}
where we have assumed $CP$ conservation, and a soft $\mathbb{Z}_2$ symmetry breaking term $m_{12}^2$.
For the neutral CP-even Higgs, with $\alpha$ as the rotation angle diagonalizing the CP-even Higgs mass matrix,
\begin{eqnarray}
\label{eq:fields}
\left(\begin{array}{c}
H\\
h
\end{array}\right) =
\left(
\begin{array}{cc}
\cos {\alpha}& \sin {\alpha}\\
-\sin {\alpha} & \cos {\alpha}
\end{array}
\right)
\,\,
\left(\begin{array}{c}
\phi_1^0\\
\phi_2^0
\end{array}\right),
\label{eq:rotation_angle}
\end{eqnarray}
In this work we set $m_H>m_h=125 \gev$, and by convention, here we set $0\leq \beta\leq \frac{\pi}{2}, 0\leq \beta-\alpha\leq \pi $.
The most general Yukawa interactions of $\Phi_{1,2}$ with the SM fermions under the $Z_2$ symmetry is
\begin{equation}\label{}
  -\mathcal{L}_{Yuk}=Y_{u}{\overline Q}_Li\sigma_2\Phi^*_u u_R^{} +Y_{d}{\overline Q}_L\Phi_dd_R^{}+Y_{e}{\overline L}_L\Phi_e e_R^{}+\text{h.c.}
\end{equation}
where $\Phi_{u,d,e}$ are either $\Phi_1$ or $\Phi_2$.
 Depending on the interactions of $\Phi_i$ coupling to the fermion sector, there are typically four types of 2HDM:

\begin{table}[h]
\centering
\begin{tabular}{c|c|c|c}
  \hline
  \hline
   		&up-type 	& down-type & leptop
  \\
  \hline
  Type-I & $\Phi_1$ &$\Phi_1$&$\Phi_1$\\
  Type-II & $\Phi_1 $ & $\Phi_2$&$\$Phi_2$\\
  Type-LS &$ \Phi_1 $&$\Phi_1$&$\Phi_2$\\
  Type-F & $\Phi_1 $&$\Phi_2$&$\Phi_1$\\
  \hline
  \hline
\end{tabular}
\caption{Interactions between fermions and Higgs doublets in four types of 2HDM.}
\end{table}

For a  review on different types of 2HDM as well as the  phenomena, see Ref.~\cite{Branco:2011iw}. \autoref{tab:yukawa_tab} is Higgs couplings to the SM fermions in the four different types of 2HDM, normalized to the corresponding SM values, for a better analysis at following sections.

\begin{table}[h]
\begin{center}
{\renewcommand\arraystretch{1.2}
\begin{tabular}{c|ccccccccc}\hline\hline
&\multicolumn{9}{c}{Normalized Higgs couplings}\\\cline{2-10}
&$\kappa_h^u$&$\kappa_h^d$&$\kappa_h^e$&$\kappa_H^u$&$\kappa_H^d$&$\kappa_H^e$&$\kappa_A^u$&$\kappa_A^d$&$\kappa_A^e$\\\hline
Type-I&$\frac{\cos\alpha}{\sin\beta}$&$\frac{\cos\alpha}{\sin\beta}$&$\frac{\cos\alpha}{\sin\beta}$&$\frac{\sin\alpha}{\sin\beta}$&$\frac{\sin\alpha}{\sin\beta}$&$\frac{\sin\alpha}{\sin\beta}$&$\cot\beta$&$-\cot\beta$&$-\cot\beta$\\\hline
Type-II&$\frac{\cos\alpha}{\sin\beta}$&$-\frac{\sin\alpha}{\cos\beta}$&$-\frac{\sin\alpha}{\cos\beta}$&$\frac{\sin\alpha}{\sin\beta}$&$\frac{\cos\alpha}{\cos\beta}$&$\frac{\cos\alpha}{\cos\beta}$&$\cot\beta$&$\tan\beta$&$\tan\beta$\\\hline
Type-LS&$\frac{\cos\alpha}{\sin\beta}$&$\frac{\cos\alpha}{\sin\beta}$&$-\frac{\sin\alpha}{\cos\beta}$&$\frac{\sin\alpha}{\sin\beta}$&$\frac{\sin\alpha}{\sin\beta}$&$\frac{\cos\alpha}{\cos\beta}$&$\cot\beta$&$-\cot\beta$&$\tan\beta$\\\hline
Type-F&$\frac{\cos\alpha}{\sin\beta}$&$-\frac{\sin\alpha}{\cos\beta}$&$\frac{\cos\alpha}{\sin\beta}$&$\frac{\sin\alpha}{\sin\beta}$&$\frac{\cos\alpha}{\cos\beta}$&$\frac{\sin\alpha}{\sin\beta}$&$\cot\beta$&$\tan\beta$&$-\cot\beta$\\\hline\hline
\end{tabular}}
\caption{Higgs couplings to the SM fermions in the four different types of 2HDM, normalized to the corresponding SM values. }
\label{tab:yukawa_tab}
\end{center}
\end{table}
In the following sections, we will take $\kappa_x=\kappa_h^x$.
For normalized SM-like Higgs gauge couplings, $V=Z,W^\pm$,
\begin{equation}
    \kappa_V\equiv\frac{g_{\rm hVV}^{\rm 2HDM}}{g_{\rm hVV}^{\rm SM}}=\sin(\beta-\alpha)
\end{equation}
with $sign(\kappa_V)=1$ by convention.

After EWSB, three Goldstone bosons are eaten by the SM gauge bosons $Z$, $W^\pm$, providing their masses. The remaining physical mass eigenstates are $h, H, A$ and $H^\pm$.  Instead of the eight parameters appearing in the Higgs potential $m_{11}^2, m_{22}^2, m_{12}^2, \lambda_{1,2,3,4,5}$, a more convenient choice of the parameters is $v, \tan\beta, \alpha, m_h, m_H, m_A, m_{H^\pm}, m_{12}^2$.

\subsection{Wrong-sign Yukawa of 2HDM}
\label{sec:wrong_sign_theory}

Taking the notations in \cite{Ferreira:2014naa}, we define,

\begin{equation}
\kappa_U \equiv \frac{\cos \alpha}{\sin \beta} =1+\cosba \cot\beta-\frac{1}{2}\cos^2(\beta-\alpha)+\mathcal{O}(\cos^2(\beta-\alpha))
\label{eq:ku-type}
\end{equation}

\begin{equation}
\kappa_D 
\equiv -\frac{\sin\alpha}{\cos\beta} 
=1- \cos (\beta-\alpha) \tan \beta-\frac{1}{2} \cos^2 (\beta-\alpha) +\mathcal{O}(\cos^2(\beta-\alpha))
\label{eq:kd-type}
\end{equation}

When $\sin(\beta-\alpha)=1$, all the SM-like Higgs boson couplings in four types will be exact same as them in SM respectively, which is the usual case called as $alignment~limit$.
These terms also can be written in the other mode,
\begin{eqnarray}
    \kappa_U &=& \sin(\beta+\alpha)+\cos (\beta+\alpha)\cot \beta
    \nonumber\\
    &=& \pm 1+\cos (\beta+\alpha)\cot \beta  \mp \frac{1}{2}\cos^2(\beta+\alpha)+\mathcal{O}(\cos^2(\beta+\alpha))
    \label{eq:ku-type-wrongsign}
\end{eqnarray}
\begin{eqnarray}
    \kappa_D &=&-\sin(\beta+\alpha)+\cos (\beta+\alpha)\tan \beta
     \nonumber\\
    &=&
    \mp 1+ \cos (\beta+\alpha) \tan \beta \pm \frac{1}{2} \cos^2 (\beta+\alpha) +\mathcal{O}(\cos^2(\beta+\alpha))
    \label{eq:kd-type-wrongsing}
\end{eqnarray}
Here we can get $\sin(\beta+\alpha) = 1, \kappa_U=-\kappa_D=1$, whilst  $\sin(\beta+\alpha) = -1, \kappa_U=-\kappa_D=-1$, which is usually called "Wrong-sign" Yukawa region in 2HDM. Through \autoref{tab:yukawa_tab}, Type-I 2HDM only has the wrong $\kappa_U=-1$ case, and other three types would have both $\kappa_U=-1$ or $\kappa_D=-1$ cases. For the gauge couplings $\kappa_V=\sin(\beta-\alpha)$, it would deviate from 1 significantly, which could be one important constraint for parameter space of the wrong-sign Yukawa region.

But even at future lepton colliders, the wrong-sign Yukawa region at tree level will be allowed as shown in \autoref{fig:LHC_tree_tbcba}, even the allowed $|\cosba|$ is less than 0.007. This situation can be changed once the loop level corrections are included,
$$\kappa_U^{\rm loop} = \kappa_U+\Delta_U^{\rm loop}\,,~~
\kappa_D^{\rm loop} = \kappa_D+\Delta_D^{\rm loop}\,.$$
$|\kappa_U^{\rm loop}|$ and $|\kappa_D^{\rm loop}|$ would not be exact 1 at same time until the decoupling effect comes. In this work, we will address $\Delta^{\rm 1-loop}$ effects to the global fit results around wrong-sign Yukawa region, with Higgs precision measurement at current LHC Run-II and future HL-LHC, CPEC.


%
%
\section{Study method}
\label{sec:study_method}
%
%
Since the discovery of 125 GeV Higgs boson at LHC Run-I, the study of Higgs sector, both the SM-like Higgs boson precision measurements and direct search of additional Higgs boson, has fruitful results. To have a complete study of wrong-sign Yukawa region of 2HDM, here we will explore its properties with both direct and indirect experimental reports at LHC Run-II.

To interpret the experimental direct search reports, we take the cross section times branching ratio $\sigma \times Br$ limits of various channels, including 
$A/H \to \mu\mu$ \cite{CMS:2016tgd, Sirunyan:2019tkw, Aaboud:2019sgt}, 
$A/H \to bb$ \cite{Sirunyan:2018taj,Aad:2019zwb}, 
$A/H \to \tau\tau$ \cite{CMS:2017epy,Sirunyan:2018zut, Aaboud:2017sjh}, 
$A/H \to tt$ \cite{Sirunyan:2019wph,Aaboud:2019roo}, 
$H \to ZZ$ \cite{Sirunyan:2018qlb, Aaboud:2017rel}, 
$H\to WW$ \cite{CMS:2016jpd, Aaboud:2017gsl}.
About the theoretical predictions in the 2HDM parameter space, we get $\sigma \times Br$ with the {\tt SusHi} package \cite{Liebler:2016ceh} for the production cross-section at NNLO level, and {\tt 2HDMC}~\cite{Eriksson:2009ws} code for Higgs decay branching ratio at tree level. 

About the indirect search, we transfer the errors of SM-like Higgs boson couplings to the constraints on the model parameters at one-loop level~\cite{Chen:2018shg}. We make a global fit by constructing the $\chi^2$ with the profile likelihood method
\begin{equation}\label{eq:chi2}
  \chi^2 = \sum_{i} \frac{(\mu_i^{\rm BSM}-\mu_i^{\rm obs})^2}{\sigma_{\mu_i}^2} .
\end{equation}
 Here $\mu_i^{\rm BSM} = \frac{(\sigma\times \text{Br})_{\rm BSM}}{(\sigma\times \text{Br})_{\rm SM}}$ for various Higgs search channels and $\sigma_{\mu_i}$ is the experimental precision on a particular channel.  
$\mu_i^{\rm BSM}$ is predicted in each specific model, depending on model parameters.   
For the LHC Run-II, the measured $\mu_i^{\rm obs}$ and corresponding $\sigma_{\mu_i}$ are given by ATLAS at 13 TeV up to 80 $fb^{-1}$ \cite{ATLAS:2019slw}. In our analyses of the future colliders, $\mu_i^{\rm obs}$ are set to be the SM value: $\mu_i^{\rm obs}=1$, assuming no deviation to the SM observables are observed. For the corresponding  $\sigma_{\mu_i}$ of the HL-LHC and CEPC, we take the precision measurements from \cite{Cepeda:2019klc,CEPCStudyGroup:2018ghi}. The future FCC-ee \cite{Abada:2019zxq} has similar performance to CEPC~\cite{Gu:2017ckc}, thus here we will only show the results with CEPC.
For one or two parameter fit, the corresponding $\Delta \chi^2=\chi^2 -\chi^2_{\rm min}$ for 95\% C.L. is 3.84 or 5.99,  respectively.

In 2HDMs, the additional Higgs sector involves several Higgs self-couplings, which are constrained by various theories considerations, such as vacuum stability,  perturbativity and unitarity. For the detailed study, we refer to the results in works~\cite{Gu:2017ckc,Chen:2018shg}.
The general idea is $-(125 \gev)^2\leq \lambvs\leq (600 \gev)^2$, and we will study inside of this region.






\section{Results at tree level}
\label{sec:results_tree}
Based on the discussion above, first we will show our study results at tree level. It includes the current LHC direct and indirect searches, as well as the indirect searches at future HL-LHC and CEPC.

\subsection{ Indirect search at LHC and future colliders}

With the global fit methods in \autoref{sec:study_method}, here we will utilize the SM-like Higgs precision measurement from LHC Run-II \cite{ATLAS:2019slw}, HL-LHC \cite{Cepeda:2019klc} and CPEC \cite{CEPCStudyGroup:2018ghi}. In details, for LHC Run-II we work with the ATLAS results ATLAS at 13 TeV up to 80 $fb^{-1}$, and for HL-LHC, we work with combined results from future ATLAS and CMS, up to 6 $ab^{-1}$. For CPEC, the latest designed luminosity is 5.6 $ab^{-1}$ at $\sqrt{S}=240 \gev$.

\begin{figure}[h]
  \centering
    \includegraphics[width=0.45\linewidth]{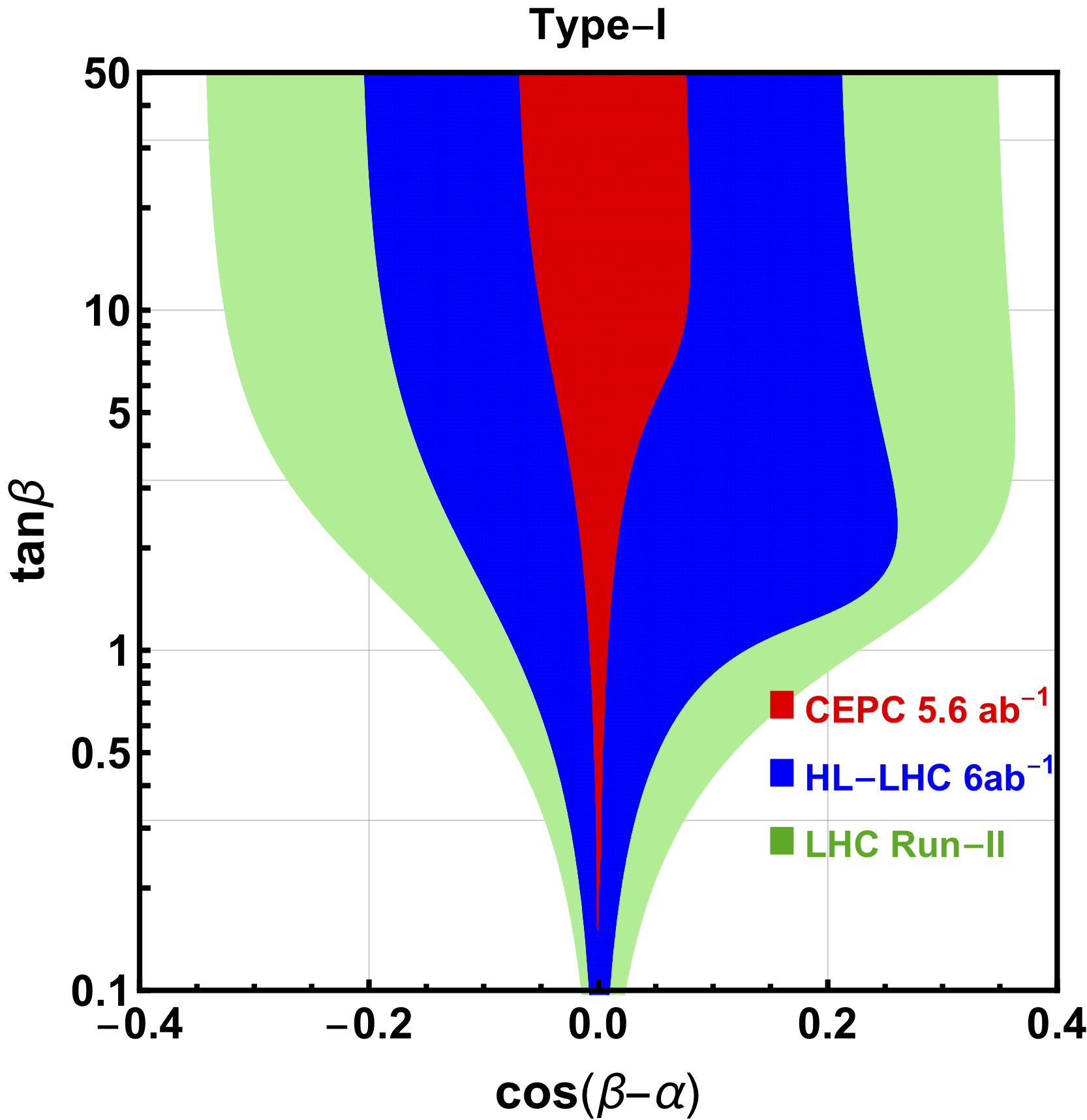}
    \includegraphics[width=0.45\linewidth]{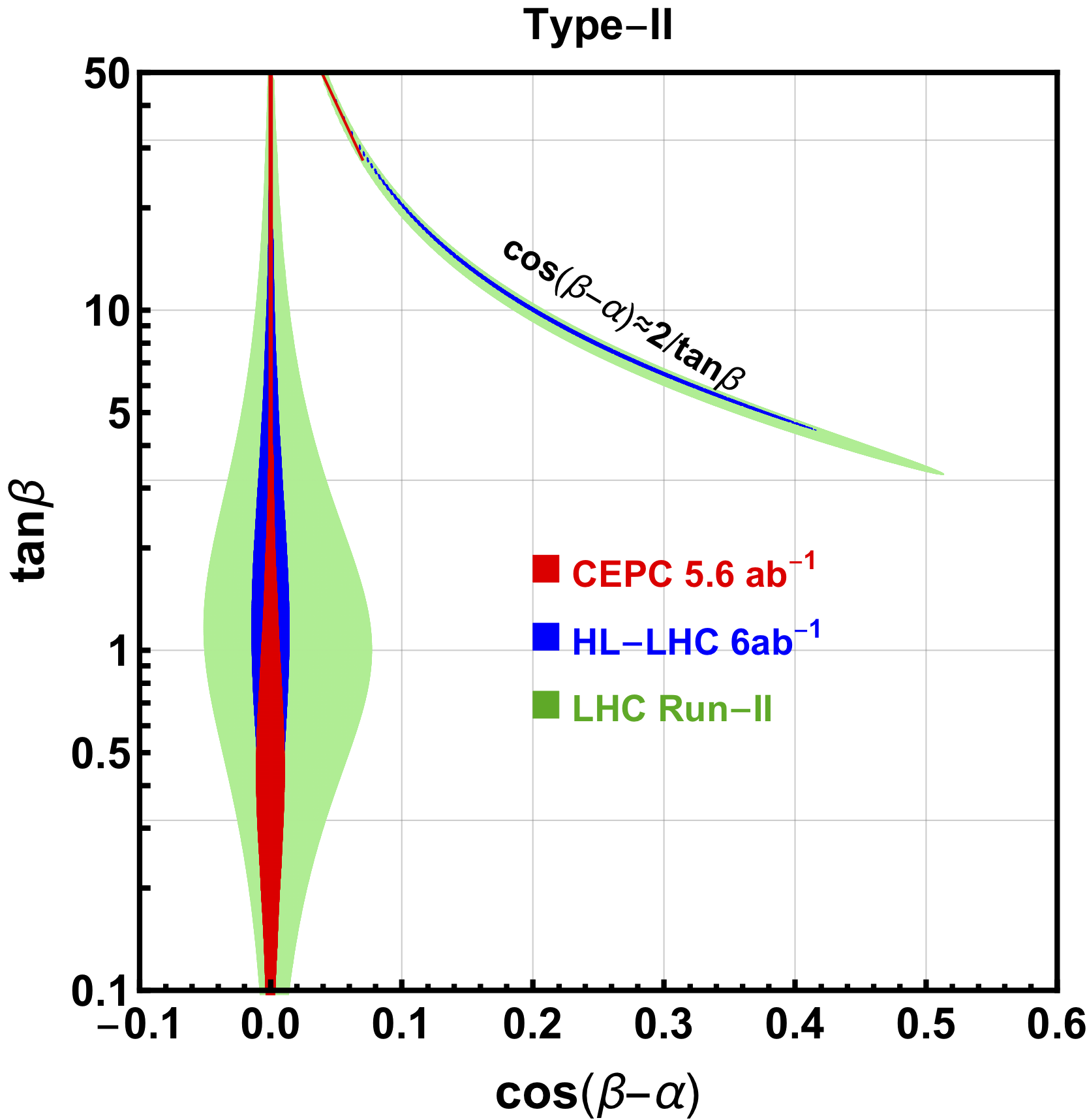}\\
    \vspace{3mm}
    \includegraphics[width=0.45\linewidth]{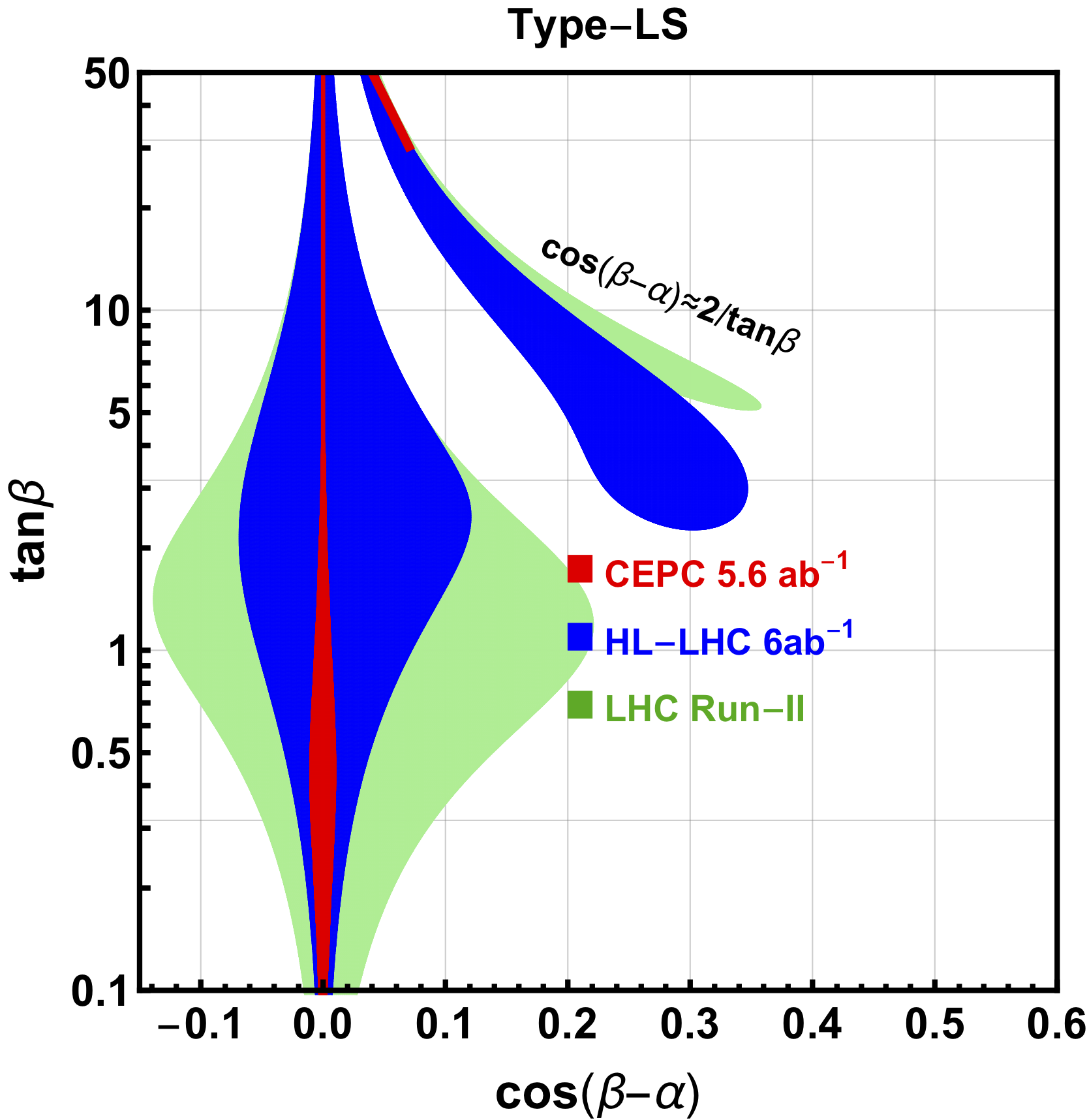}
    \includegraphics[width=0.45\linewidth]{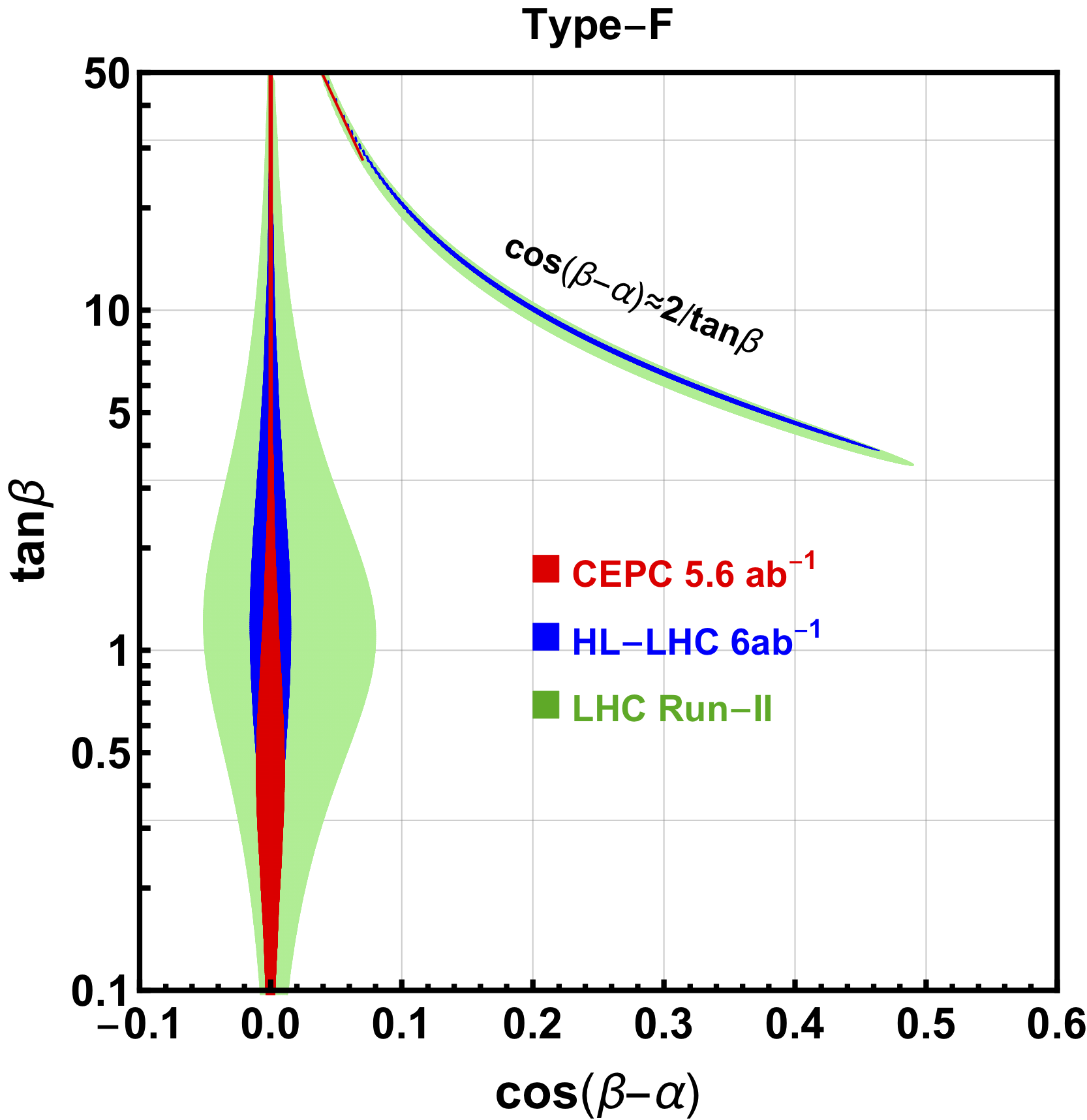}
  \caption{The allowed region in the  plane of $\tan \beta$ - $\cos(\beta-\alpha)$ at 95\% C.L. for the four types of 2HDM, given LHC Run-II (green), HL-LHC (blue) and CEPC (red) Higgs precision measurements.  For future measurements, we assume that the measurements agree with SM predictions.  The special  ``arm" regions for the Type-II, L and F are the wrong-sign Yukawa regions as discussed in \autoref{sec:wrong_sign_theory}}
  \label{fig:tanb_cba_tree}
\end{figure}
We give our global fit results in \autoref{fig:tanb_cba_tree}, the allowed region in the  plane of $\tan \beta$ - $\cos(\beta-\alpha)$ at 95\% C.L. for the four types of 2HDM, given LHC Run-II (green), HL-LHC (blue) and CEPC (red) Higgs precision measurements. For the Type-I 2HDM, all the SM-like Higgs fermion couplings are $\kappa_U$ type in \autoref{eq:ku-type} with $\cot\beta$-enhanced corrections when deviates from $alignment~limit$  $\cosba=0$. Thus at large $\tanb$ region, the Yukawa couplings would not contribute much in constraining the parameter space, and the main restriction is from gauge couplings. The detailed values are displayed in \autoref{tab:tree_results}.

\begin{table}[h]
\begin{center}
\begin{tabular}{|c|c|c|c|}
\hline
Type                    & LHC Run-II
&HL-LHC (ATLAS+CMS)     &CEPC
\\ \hline
Type-I $\tan\beta \gtrsim 5$    & 0.38       & 0.2                        &0.08
\\ \hline
Type-II $\tan\beta\sim1$   & 0.08            & 0.015                         &0.01
\\ \hline
Type-L $\tan\beta\sim1$  & 0.22       & 0.12                               &   0.011
\\ \hline
Type-F $\tan\beta\sim1$   & 0.08       & 0.015                              & 0.012
\\ \hline
\end{tabular}
\end{center}
\caption{Apart for the wrong-sign region, the maximally allowed $|\cos(\beta-\alpha)|$ range at 95\% C.L. given LHC Run-II, HL-LHC and CEPC Higgs precision measurements.    }
\label{tab:tree_results}
\end{table}

For the other three types, they include both $\kappa_U$ and $\kappa_D$ type Yukawa couplings, as a result both large and small $\tanb$ are strongly constrained apart for the wrong-sign Yukawa regions. The relevant the maximally allowed $|\cos(\beta-\alpha)|$ ranges are also shown in~\autoref{tab:tree_results}. We also note the Type-LS is less restricted at small $\tanb$ compared to Type-II and Type-F, because only lepton couplings of Type-LS have $\kappa_D$ type and the precisions of $\delta \kappa_{b} $ is better than  $\delta \kappa_{\tau} $, for example in CPEC, $\delta \kappa_{b} =1.3\%, \delta \kappa_\tau=1.5\% $.

%

\subsection{ Wrong-sign region and disappeared up-type}

Also shown in~\autoref{fig:tanb_cba_tree}, there are  regions with large $\cosba$, the upper right "arm" region. From \autoref{eq:ku-type-wrongsign} and \autoref{eq:kd-type-wrongsing}, even $\cosba \nRightarrow   0$ there are still allowed regions to get $|\kappa_{U,D}|=1$, the so called wrong-sign Yukawa region of 2HDM.

As discussed in~\cite{Ferreira:2014naa}, the $\kappa_D$-type wrong-sign Yukawa in~\autoref{eq:kd-type-wrongsing} only occurs at $\tanb>1$. For the exact $\kappa_U=-\kappa_D=1$, $\sin(\beta-\alpha)=-\cos2\beta$, and at large $\tanb$, we have 
\begin{equation}
    \cosba=2/\tanb\,.
    \label{eq:wrong-sign-tree-right}
\end{equation}
 Thus even at CEPC, where we will have $\delta \kappa_Z = |1-\sin(\beta-\alpha)| \leq 0.25\%$, the wrong-sign Yukawa is still allowed around $\cosba\approx 2/\tanb$ for $\cosba<0.07$ at tree level.

$\kappa_U$-type wrong-sign Yukawa in~\autoref{eq:kd-type-wrongsing} only occurs at $\tanb<1$. For the exact $\kappa_D=-\kappa_U=1$, $\sin(\beta-\alpha)=\cos2\beta$, and at small $\tanb$,
\begin{equation}
    \cosba=-2\tanb\,.
    \label{eq:wrong-sign-tree-left}
\end{equation}

Usually $\kappa_U$ and $\kappa_D$ are estimated in the form of $\kappa_{U,D}^2$, except for if there is any interference.  The two sensitive parameters~\cite{Cepeda:2019klc} are 

\begin{align}
&\kappa_\gamma = (1.59 \kappa_W^2 -0.67 \kappa_t \kappa_W+ 0.071 \kappa_t^2...)^{0.5}\,,\label{eq:kgamma_tree}\\
&\kappa_g = (1.11 \kappa_t^2 -0.12 \kappa_t \kappa_b+ 0.01 \kappa_b ^2...)^{0.5}\,.\label{eq:kg_tree}
\end{align}
Here \autoref{eq:kgamma_tree} and \autoref{eq:kg_tree} tell us the sign of $\kappa_b$ does not make an important enough difference to $\chi^2(\kappa_b\to 1)$ and $\chi^2(\kappa_b \to -1)$ through the global fit method~\autoref{eq:chi2} at tree level~\cite{Ferreira:2014naa}, while the sign of $\kappa_t$ makes an important difference to both $\kappa_\gamma, \kappa_g$. For $\kappa_U$-type wrong-sign region, corrected $\kappa_{\gamma\gamma}$ deviated from SM values too large to be excluded.
%

%
\subsection{Current LHC direct search}
\label{sec:LHC_tree}

After the indirect searches, here we will take the Type-II 2HDM as an example to compare with the direct LHC searches, and to explore the combined constraint ability to the wrong-sign Yukawa region.

\begin{figure}[ht]
  \centering
  \includegraphics[width=0.3\linewidth]{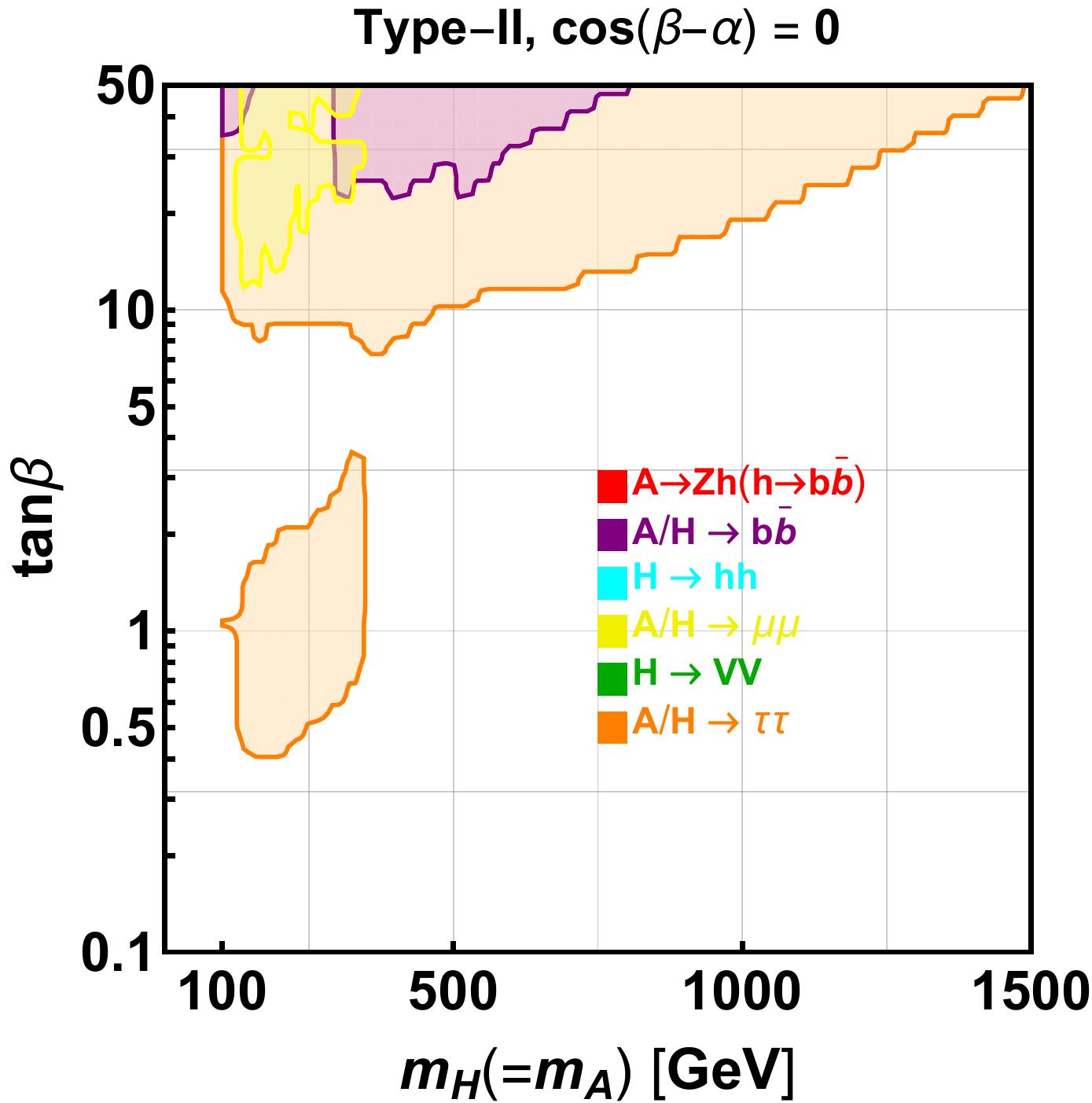}
    \includegraphics[width=0.3\linewidth]{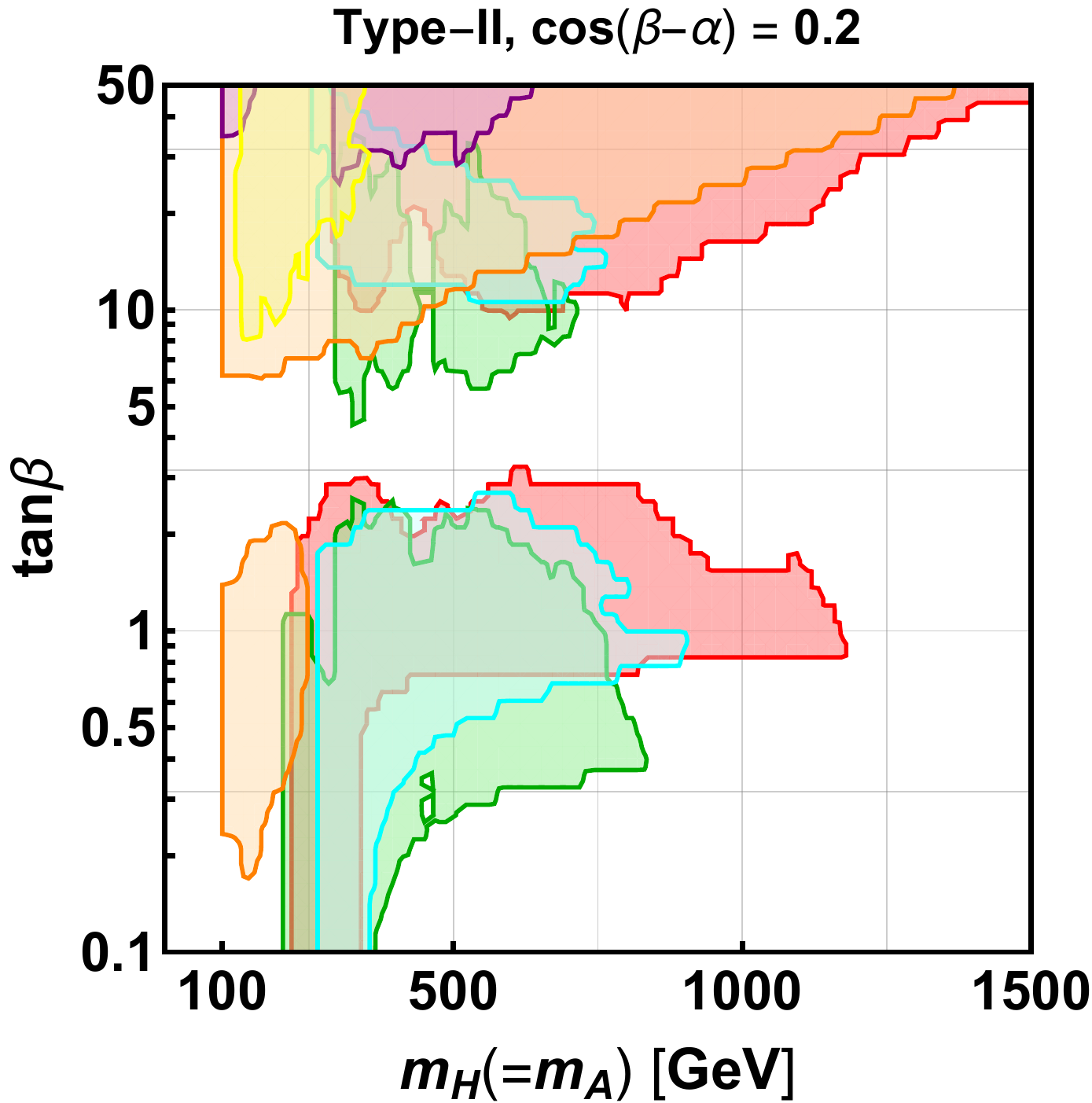}
    \includegraphics[width=0.3\linewidth]{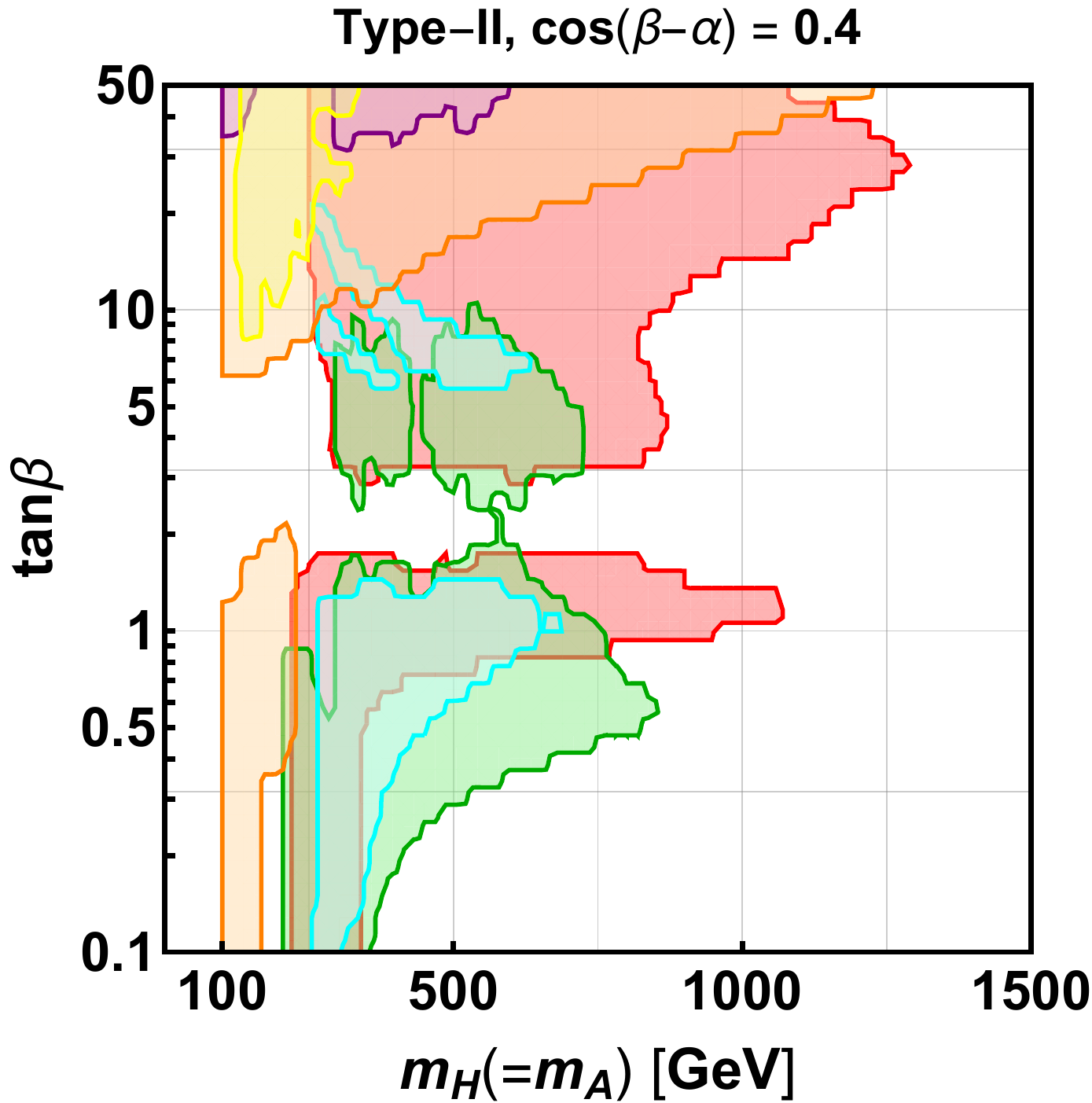}
  \caption{Interpretation results of excluded region in the plane $m_{H/A} - \tanb$ with latest LHC direct search limits, including $A\to Zh (h\to b\bar b)$ (red), $A/H \to b \bar b$ (purple), $H \to hh$ (cyan),  $A/H \to \mu^+\mu^-$ (yellow), $H \to VV$ (green),  $A/H \to \tau^+\tau^-$ (orange). Here the benchmark parameter is degenerate heavy Higgs mass $m_A=m_H=m_{H^+}, ~\cosba=$ 0 (left), 0.2 (middle), 0.4 (right), and $m_H^2 = m_{12}^2/s_\beta c_\beta$.  }
  \label{fig:LHC_tree_mphitanb}
\end{figure}

As shown in \autoref{fig:LHC_tree_mphitanb}, the excluded region by current LHC direct search in the plane $m_{H/A} - \tanb$,  including $A\to Zh (h\to b\bar b)$ (red), $A/H \to b \bar b$ (purple), $H \to hh$ (cyan),  $A/H \to \mu^+\mu^-$ (yellow), $H \to VV$ (green),  $A/H \to \tau^+\tau^-$ (orange) respectively. Based one \autoref{fig:tanb_cba_tree}, to study the the wrong-sign Yukawa region, we take the benchmark parameter $\cosba=$ 0 (left), 0.2 (middle) , and 0.4 (right), with degenerate heavy Higgs mass $m_A=m_H=m_{H^+}$, $m_H^2 = m_{12}^2/s_\beta c_\beta$.

In the left panel of \autoref{fig:LHC_tree_mphitanb}, only $H/A\to f\bar f$ channels have constraint since $Hhh, HVV, AhZ$ couplings at tree level are proportional to $\cosba$. Generally the region $m_A \in (130, 750)$, $\tanb>10$ is excluded by $\tau\tau$ decay channel, and for larger heavy Higgs mass, the excluded $\tanb$ limit will be larger, to limitless around 1.5 TeV. Also a small region $m_A \in (130, 2m_t), ~\tanb\in(0.5,2)$ is excluded by $A/H \to \tau\tau$. 
For middle and right panels of \autoref{fig:LHC_tree_mphitanb}, all channels here would make a difference with non-zero $\cosba$. For $\cosba=0.2$, at large $\tanb$ the regions of $m_A < 700, \tanb>5$, $m_A < 800, \tanb>10$ are excluded. Similarly the restriction ability goes down until 1.5 TeV. At small $\tanb$ region, $m_A < 750, \tanb<0.3$ is strongly constrained. The excluded region can reach 1.2 TeV for $\tanb \in (0.9,2)$. For larger $\cosba=0.4$, when $m_A < 800$, $\tanb>3$ are strongly constrained since the more powerful $A\to Zh$ channel. This channel gets larger decay rates with larger $\cosba$. But it can only reach 1.4 TeV around $\tanb=30$. The excluded region of $\cosba=0.4$ at small $\tanb$ region is similar as $\cosba=0.2$.

The strong constraints at large $\tanb$ and non-zero $\cosba$ can contribute to exclude the wrong-sign Yukawa region. To have a more straightforward idea, we will compare the direct and indirect searches in the plane $\cosba-\tanb$.

\begin{figure}[h]
  \centering
    \includegraphics[width=0.3\linewidth]{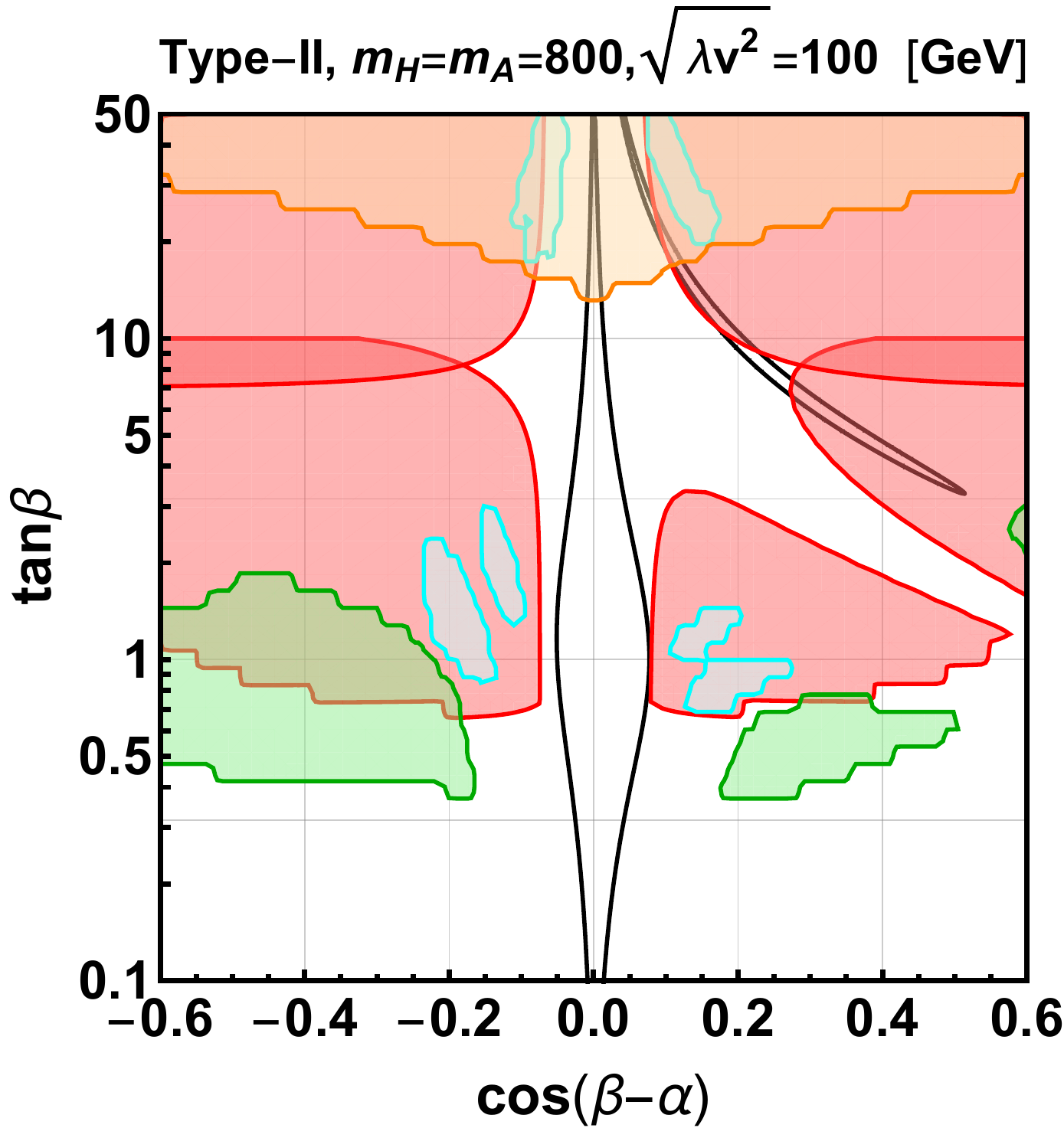}
     \includegraphics[width=0.3\linewidth]{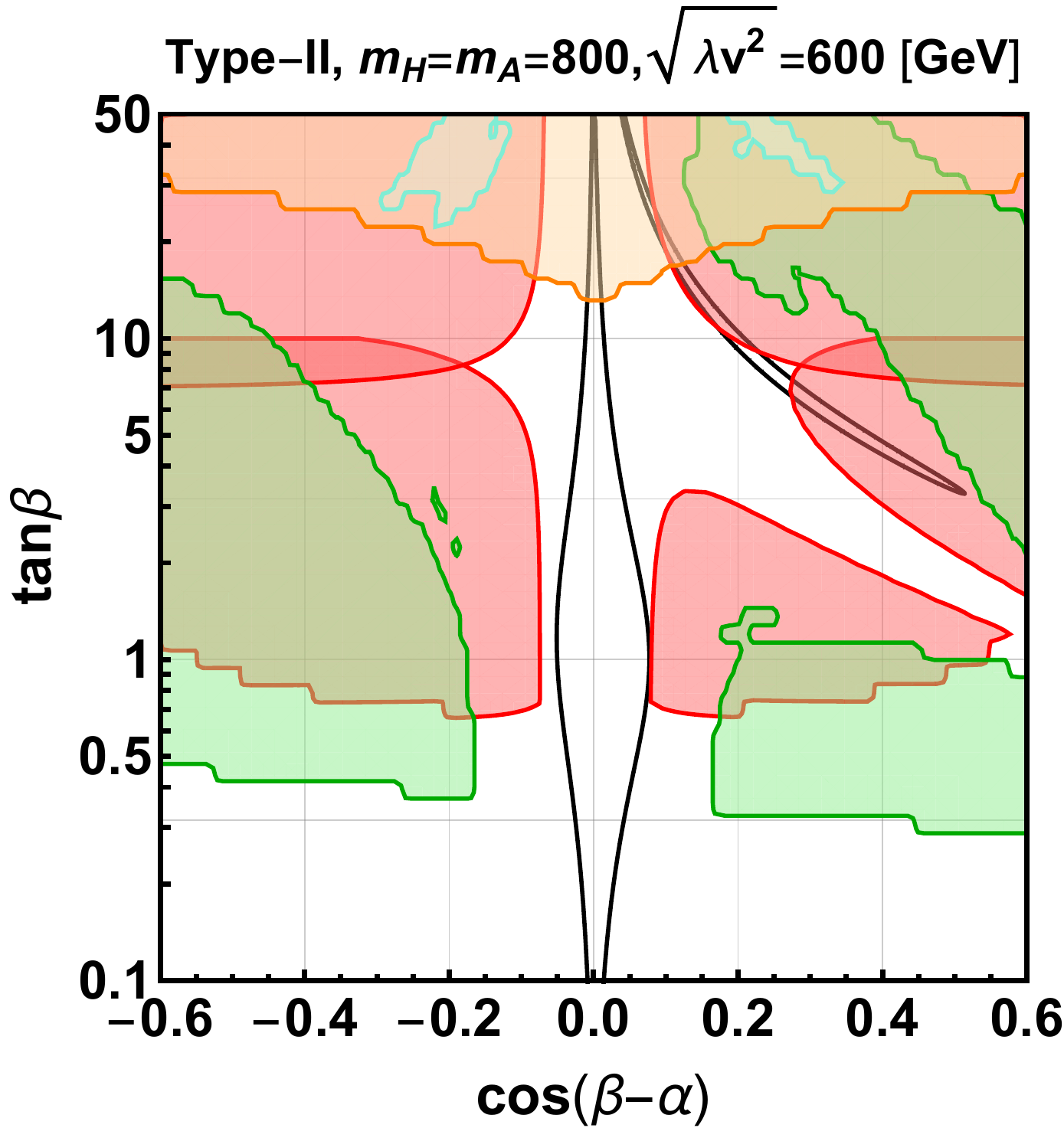}
    \includegraphics[width=0.3\linewidth]{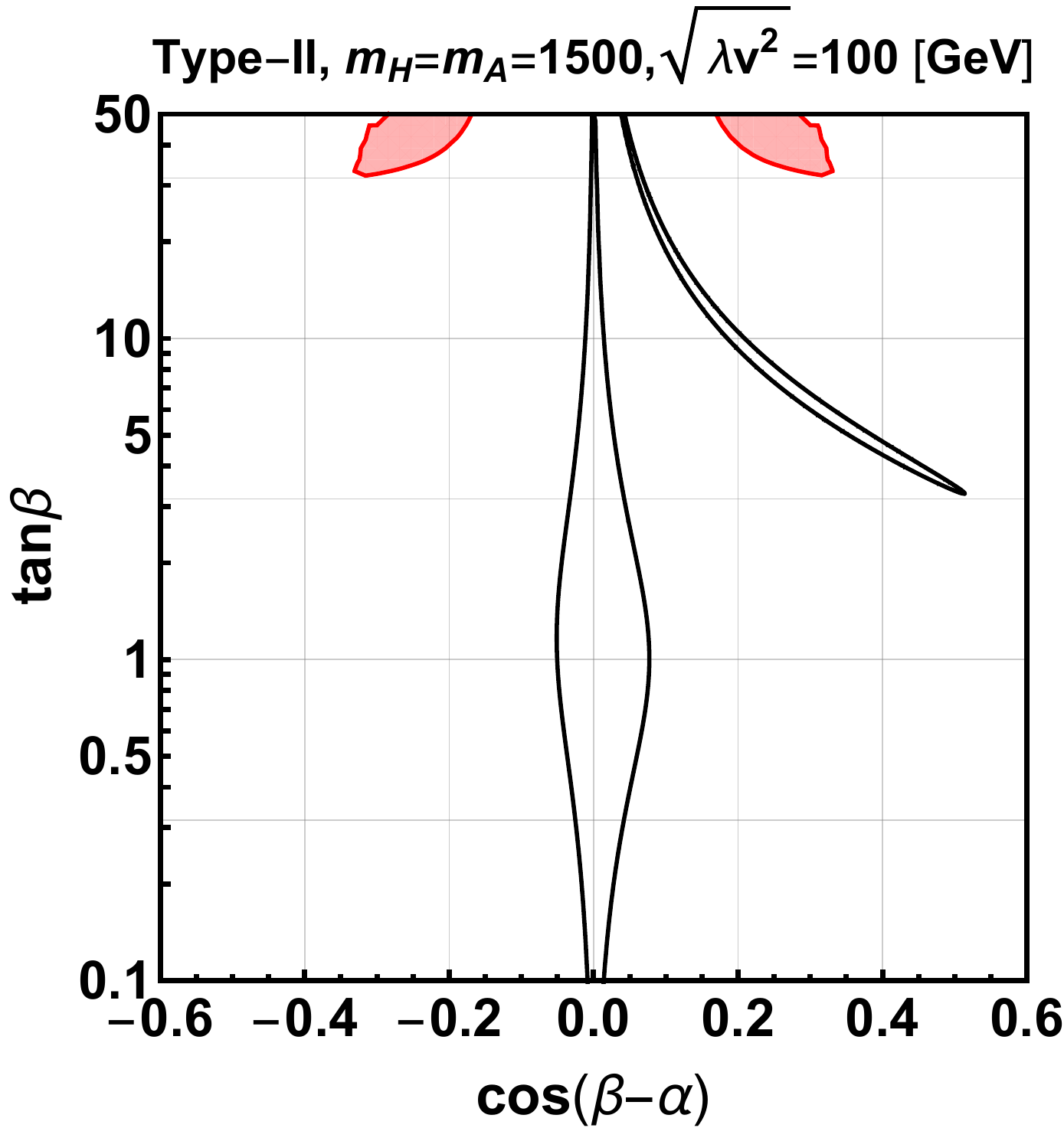}
  \caption{ 
 Here we will show the excluded region from indirect Higgs precision measurements and  direct search results in the plane $\tan\beta$ - $\cos(\beta-\alpha)$ for Type II 2HDM. The indirect come from ATLAS at 13 TeV~\cite{ATLAS:2019slw}, and the direct seaches includs $A\to HZ$ (blue), $A \to Zh$ (red), $A/H \to \tau\tau$ (orange), $A/H \to bb$ (purple) and $A/H \to \mu\mu$ (yellow), to compare different channel constraining ability. Here we choose benchmark parameters $m_A=m_H=m_{H^+} = $ 800 GeV (left and middle), 1500 GeV (right) and $ \lambvs \equiv \sqrt {(m_H^2 - m_{12}^2/s_\beta c_\beta)} = 100$ GeV (left and right), 600 GeV (middle).
The current precision measurements results are shown by solid lines. 
Generally the central region between the two lines around $\cosba=0$ are allowed except for the "arm" of Type-II, the wrong-sign Yukawa region as discussed detailed in \autoref{fig:tanb_cba_tree}.  }
  \label{fig:LHC_tree_tbcba}
\end{figure}

As in \autoref{fig:LHC_tree_tbcba}, here we choose benchmark parameters $m_A=m_H=m_{H^+} = $ 800 GeV (left and middle), 1500 GeV (right) and $ \lambvs \equiv \sqrt (m_H^2 - m_{12}^2/s_\beta c_\beta) = 100$ GeV (left and right), 600 GeV (middle), to discuss the combine the constraint from indirect Higgs precision measurement and direct heavy Higgs searches at current LHC Run-II. The details about the experimental reports are same as \autoref{fig:tanb_cba_tree} and \autoref{fig:LHC_tree_mphitanb}.
At left panel with $m_A= $ 800 GeV, $\lambvs = 100$ GeV , the wrong-sign region at large $\tanb>20$ is totally covered by $A/H \to \tau\tau$ channel, and at small $\tanb$ region, it is strongly constrained by $A\to Zh$ channel. The small allowed region is around $8<\tanb<10, 0.2<\cosba<0.3$. At small $\tanb$ region, LHC direct searches give weak constraints resulting from too wide $\Gamma_{A/H}$ and current searches are not valid in this region. Compared the middle panel with $\lambvs = 600$ GeV, the general results around wrong-sign Yukawa region are quite similar. This  tells us the independence on $\lambvs$ or $m_{12}$ in the considered regions. For the right panel with $m_A= $ 1500 GeV, $\lambvs = 100$ GeV , the LHC direct search can nearly give no constraints there, which is also shown in \autoref{fig:LHC_tree_mphitanb}. Also from middle and right panels of \autoref{fig:LHC_tree_mphitanb}, where the LHC direct search constraints are similar for $m_A<800$ GeV and large $\tanb$ region, we can say the wrong-sign region with $m_A<800$ GeV are strongly constrained by the combined indirect and direct searches at tree level.

%


\section{Results at one-loop level}
\label{sec:results_loop}
From last section, the combined indirect and direct searches at current LHC can give strong constraints on wrong-sign Yukawa region for $m_A<800 \gev$ while for large heavy Higgs mass such $m_A=1500 \gev$, direct searches nearly has no restrictions. The conclusion will be modified to a large extent when including the loop-level corrections to Higgs precision measurement study~\cite{Gu:2017ckc,Chen:2018shg}.
\subsection{Loop effects in $\cosba-\tanb$ plane}
To explore loop effects on the wrong-sign Yukawa region, here we first analyze the individual Higgs couplings cosntraints in details with Type-II 2HDM. In~\cite{Gu:2017ckc,Chen:2018shg}, we have detailed studies about the normal Yukawa regions around $\cosba=0$, and the studies method here are similar, thus here we only display the wrong-sign regions.
\begin{figure}[h]
  \centering
\includegraphics[width=0.49\linewidth]{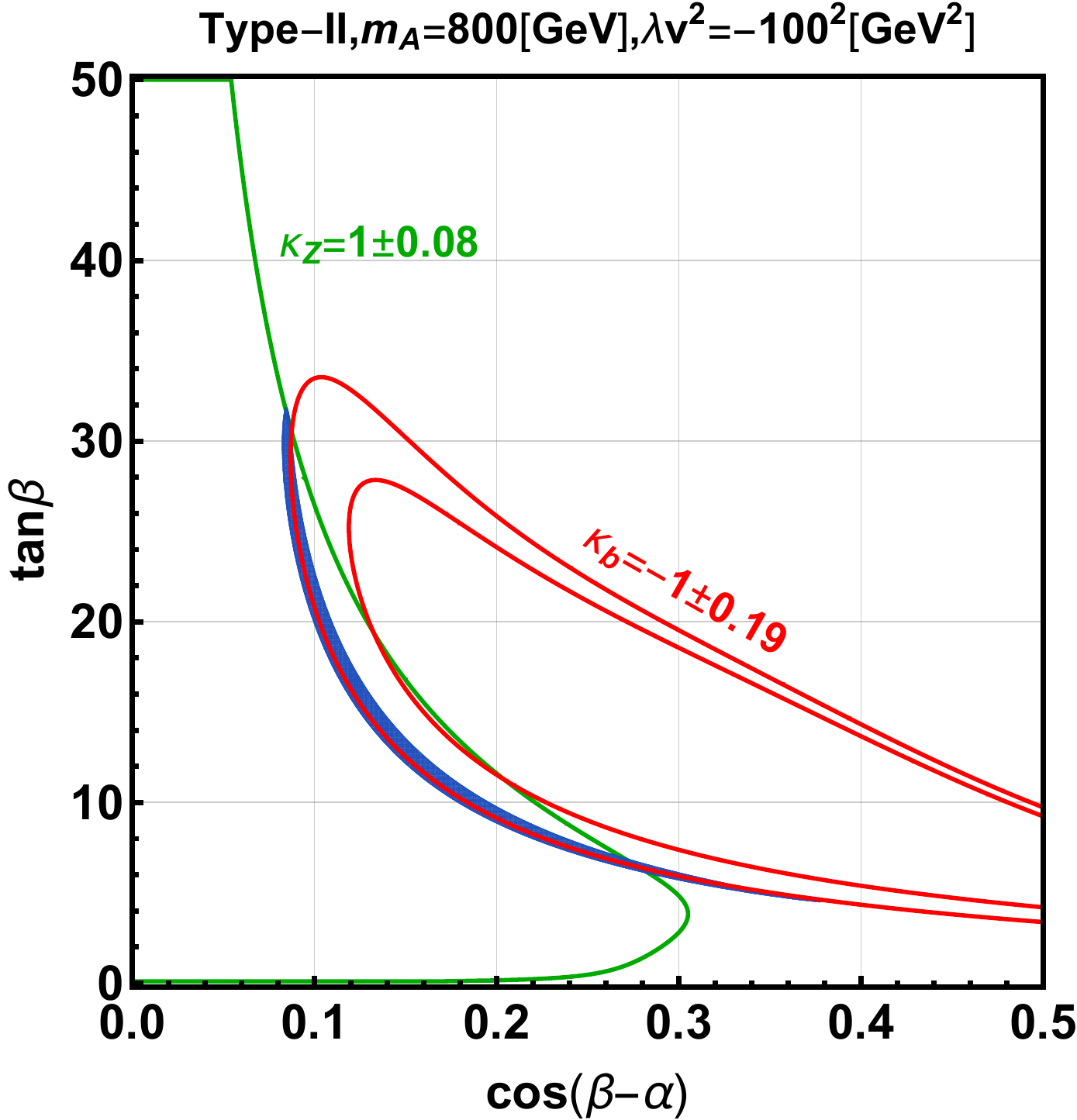}
\includegraphics[width=0.49\linewidth]{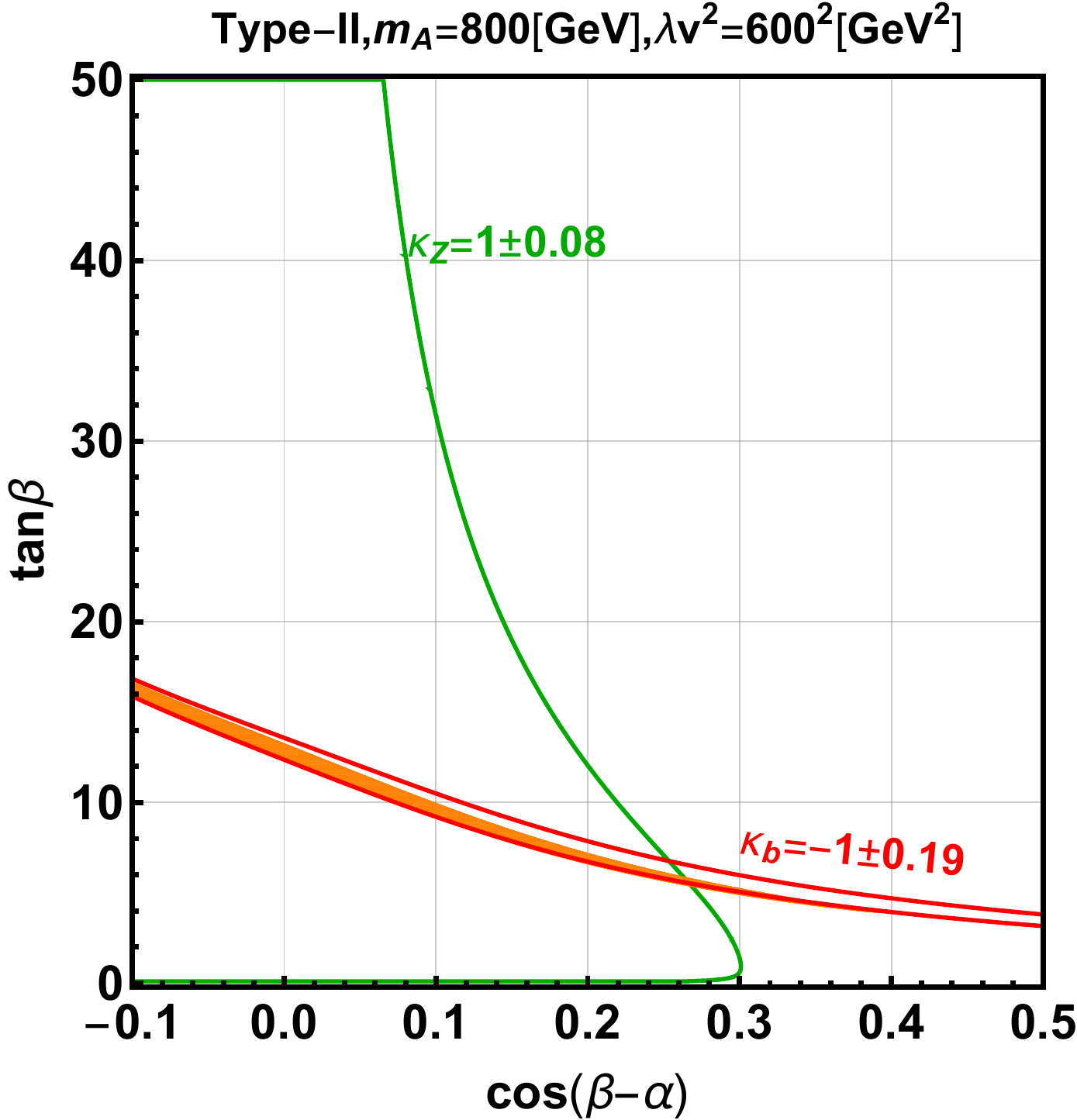}
 \caption{ The allowed region in the  plane of $\tan \beta$ - $\cos(\beta-\alpha)$ at 95\% C.L. for Type-II 2HDM, given LHC Run-II Higgs precision measurements at one-loop level. Here we only show the special  ``arm" region , the wrong-sign Yukawa regions as discussed in \autoref{sec:wrong_sign_theory} and \autoref{fig:tanb_cba_tree}. Here we take the benchmark parameters $m_A=m_H=m_{H^\pm}=800$ GeV, $\lambda v^2=-100^2$ (left), $600^2$ (right) $\gev^2$. We also show the current precision $\delta \kappa_b = \pm 0.19$ and $\delta \kappa_Z = \pm 0.08$ with red and green lines respectively.
 }
  \label{fig:tbcba_loopana}
\end{figure}

As the \autoref{fig:tbcba_loopana}, we show the allowed wrong-sign Yukawa region in the  plane of $\tan \beta$ - $\cos(\beta-\alpha)$ at 95\% C.L. for Type-II 2HDM, given LHC Run-II Higgs precision measurements at one-loop level. The benchmark parameters in the left panel is $m_A=m_H=m_{H^\pm}=800$ GeV, $\lambda v^2=-100^2$. The blue region is allowed at one-loop level, and the red and green lines are for $\delta \kappa_b = \pm 0.19$ and $\delta \kappa_Z = \pm 0.08$ taken from current LHC reports\cite{Cepeda:2019klc}. The left region of the green line is allowed, and the region between two red lines is allowed by one-loop $hbb$ coupling. The one-loop $hbb$ coupling at large $\tanb$ is largely corrected and shifted to right. Thus the combined $hZZ, hbb$ coupling would restrict $\cosba\in(0,08,0.038)$ for the wrong-sing Yukawa region. For $\lambvs=600 \gev$ at right panel, $hbb$ is strongly shifted to left, resulting to strongly shifted wrong-sign region represented by orange shadow. Generally the strong corrections comes from Higgs field renormalization, and we have detailed analysis at~~\cite{Gu:2017ckc,Chen:2018shg}.

\begin{figure}[h]
  \centering
\includegraphics[width=0.49\linewidth]{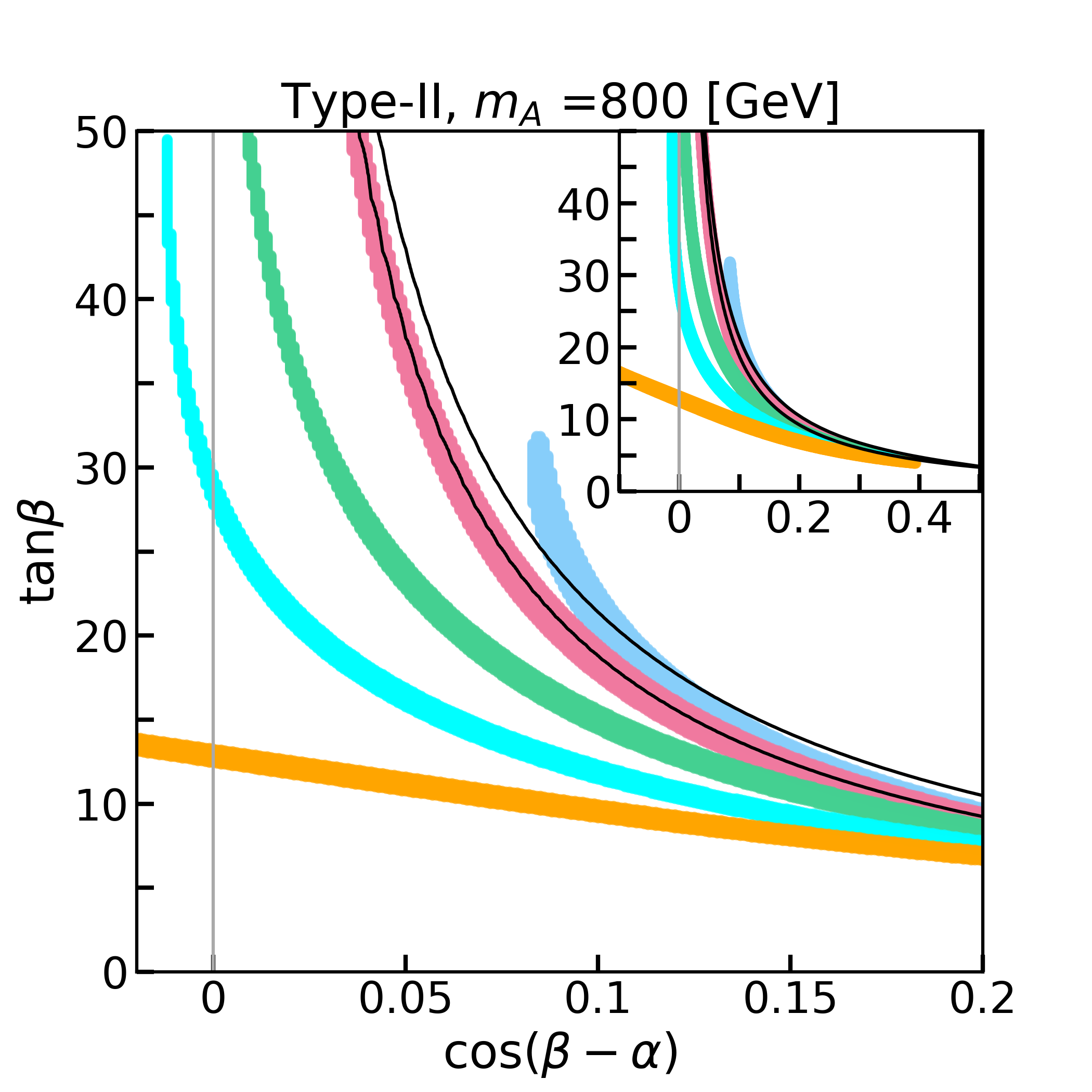}
\includegraphics[width=0.49\linewidth]{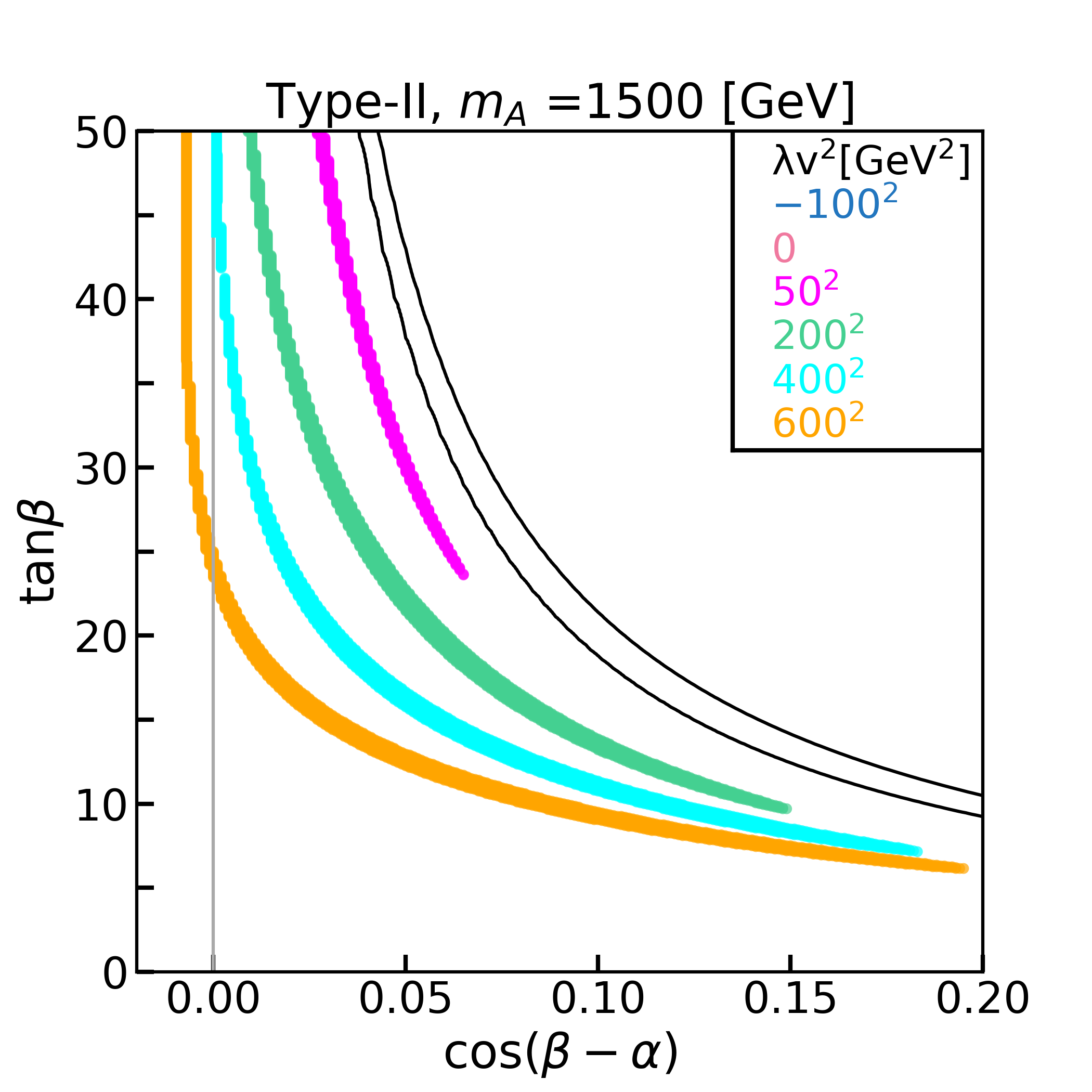}
 \caption{ The summarized allowed wrogn-sign Yukawa region in the  plane of $\tan \beta$ - $\cos(\beta-\alpha)$ at 95\% C.L. for Type-II 2HDM, given LHC Run-II Higgs precision measurements at one-loop level. Here we take the benchmark parameters $m_A=m_H=m_{H^\pm}=800$ GeV (left) and 1500 GeV (right). The diffenet colorful regions are for $\lambda v^2=-100^2$ (blue), 0 (light red), $50^2$ (magenta), $200^2$ (green), $400^2$ (cyan) and $600^2$ (orange) $\gev^2$. We also show the allowed wrong-sign Yukawa region at tree level with black solid lines. For $m_A=800 \gev$, we show the larger allowed region in the subplot, upper right corner of the left panel. }
  \label{fig:tbcba_loop_sum}
\end{figure}
In \autoref{fig:tbcba_loop_sum}, based on the analysis in \autoref{fig:tbcba_loopana}, we show the allowed wrogn-sign Yukawa region of various $\lambda v^2$ values in the  plane of $\tan \beta$ - $\cos(\beta-\alpha)$ at 95\% C.L. for Type-II 2HDM, given LHC Run-II Higgs precision measurements at one-loop level. 
In the left panel, we work with the benchmark parameters $m_A=m_H=m_{H^\pm}=800$ GeV and  $\lambda v^2=-100^2, 0, 50^2, 200^2, 400^2, 600^2 \gev^2$ displayed by blue, light red, magenta, green and cyan regions respectively.  In the main region of left panel, we show $\cosba \in (-0.02, 0.2)$ to have a clear performance. The larger region is in the subplot, and  $\cosba \in (-0.02, 0.4)$ at one-loop level. The allowed wrong-sign Yukawa region at tree level is displayed by black solid lines. Here the light red region for $\lambda v^2=0$ is the most similar one to the tree level region, and regions of smaller $\lambda v^2$ would locate at right while  regions of larger $\lambda v^2$ would locate at left. Given the analysis at \autoref{fig:LHC_tree_tbcba}, we can conclude that $\lambda v^2 <0$ will be more constrained and $\lambda v^2 >0$ will be less constrained under the combined constraints for $m_A=800 \gev$. For $m_A=1500 \gev$, region of $\lambda v^2\leq 0$ is totally excluded, and for large $\lambda v^2$ the allowed region is shifted to right of black lines and large $\cosba$ regions are also constrained. In general under the combined constraints, it is similar as $m_A=800 \gev$, $\lambda v^2 >50 \gev^2$ will be less constrained. In one word, we can conclude that with loop corrections, small $\lambda v^2$ will be more constrained  compared to tree level results,  while large $\lambda v^2$ is less constrained.
%

%
\subsection{Loop effects in $m_\Phi - m_{12}$ plane}
%
Based on \autoref{eq:wrong-sign-tree-right}, the wrong-sign Yukawa region has a simple relationship $\cosba=2/\tanb$ when  $\tanb \gg 1$ at tree level. Here we would test this relationship at one-loop level to explore radioactive effects.

\begin{figure}[h]
  \centering
\includegraphics[width=0.49\linewidth]{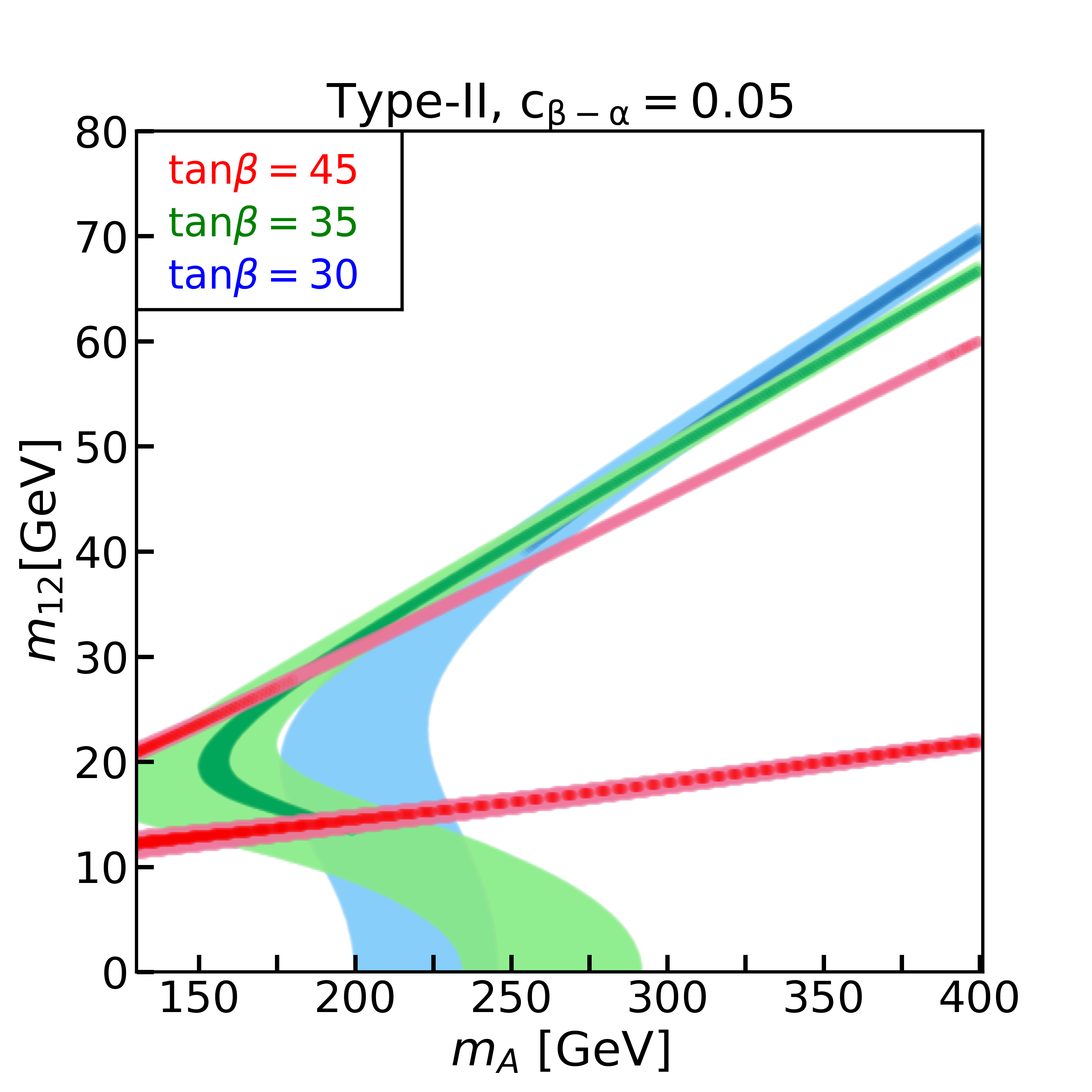}
\includegraphics[width=0.49\linewidth]{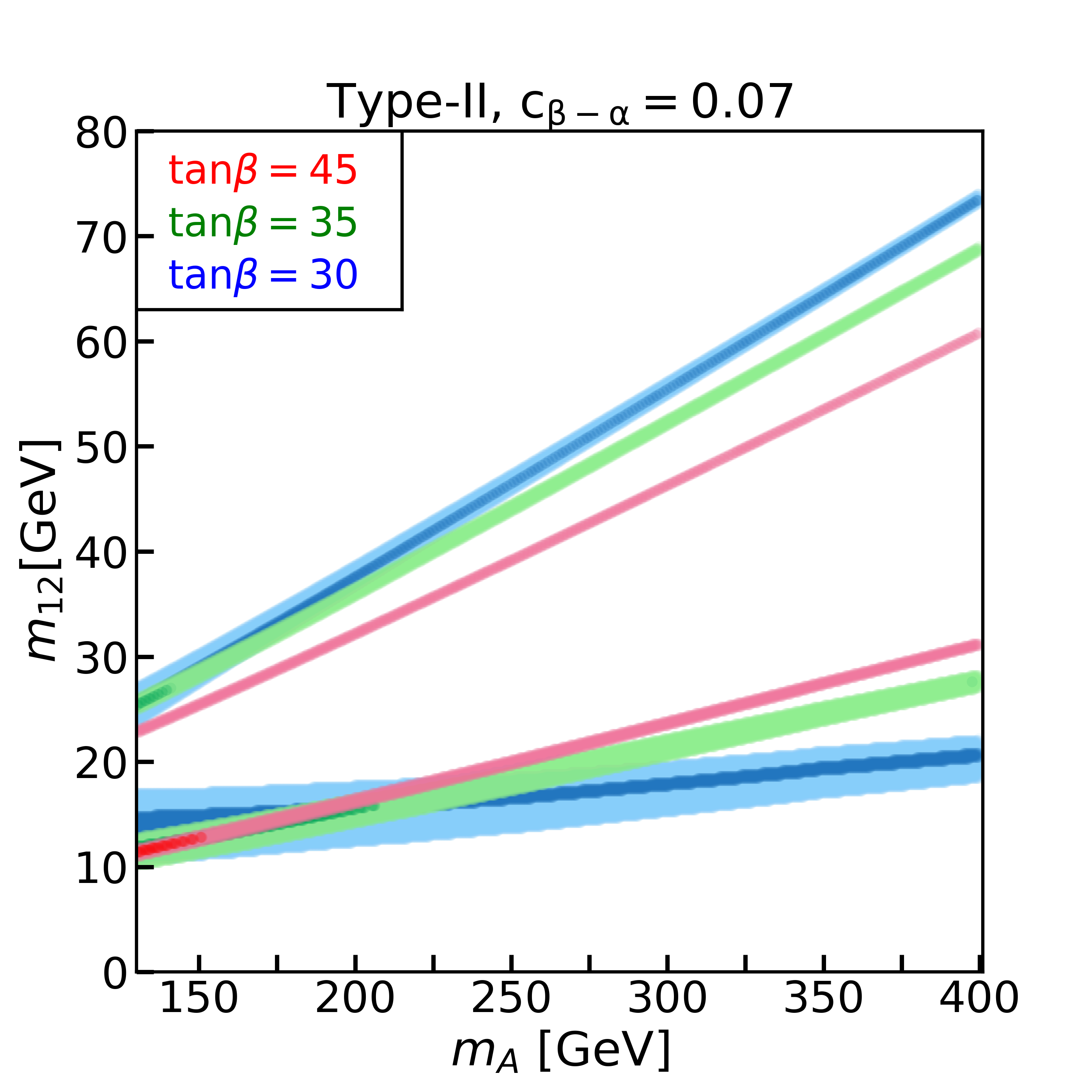}
 \caption{ The allowed region in the  plane of $m_A$ - $m_{12}$ at 95\% C.L. for Type-II 2HDM, given LHC Run-II (light color) and HL-LHC (dark color) Higgs precision measurements at one-loop level. Here we take the benchmark parameters $m_A=m_H=m_{H^\pm}$, $\cosba=$ 0.05 (left), 0.07 (right), $\tanb$ = 30 (blue), 35 (green), 45 (red) to test $\cosba=2/\tanb$. Generally for a pair of fixed $\cosba$ and $\tanb$, the allowed wrong-sign Yukawa region tends to get a linear relationship between $m_A$-$m_{12}$, until $m_A $ approaches around 2 TeV to be excluded with current LHC Run-II data. 
 }
 \label{fig:mphim12_loop}
\end{figure}
In \autoref{fig:mphim12_loop}, performing the global fit at 95\% C.L. for Type-II 2HDM, we show the allowed region in the  plane of $m_A$ - $m_{12}$ after including the loop corrections to SM-like Higgs couplings. For the benchmark parameters, we still take degenerate heavy Higgs mass $m_A=m_H=m_{H^\pm}$. In order to to meet $\tanb\gg 1$, we choose $\cosba=$ 0.05 (left), 0.07 (right), $\tanb$ = 30 (blue), 35 (green), 45 (red) to test $\cosba=2/\tanb$, which describes the wrong-sign region precisely at tree level. The global fit results with current LHC and future HL-LHC Higgs precision measurements are displayed with light and dark colors respectively. With the future CEPC reports, the allowed region is strongly constrained, and since the best $\chi^2$ is larger than 100 for these benchmark points, we would not show them here.

Generally for a pair of fixed $\cosba$ and $\tanb$, the allowed wrong-sign Yukawa regions at one-loop level are divided into two parts based on $\tanb=20$.
For $\tanb>20$, the region tends to have $\lambvs=m_A^2-m_{12}^2(1+\tan^2\beta)/\tanb \approx 0$, where the loop correction is small and the wrong-sign region is kept as the tree level. This relationship can last to $m_A$ approaching 2 TeV. Larger region is excluded.
For $\tanb<20$, the small corrections comes from the complex cancellations between the different parts during the renormalization process. 

%
%
\section{Conclusions}
\label{sec:con}

Since the discovery of SM-like Higgs boson at LHC Run-I, exploring its properties especially Higgs couplings become a promising method to study new physics. In the framework of 2HDM, this work focuses on testing the so-called wrong-sign Yukawa region up to one-loop level with both indirect and direct searches at current LHC. For the direct searches, we constrain the parameter space with various heavy Higgs decays, $A/H \to f\bar f, VV, Vh, hh$ at tree level. For the indirect searches, we perform the global fit with current LHC, future HL-LHC and CEPC Higgs precision measurements up to one-loop level.

Generally as shown in \autoref{fig:tanb_cba_tree}-\autoref{fig:LHC_tree_tbcba}, for degenerate heavy Higgs mass $m_A=m_H=m_{H^\pm}<800$ GeV
the wrong-sign Yukawa regions are excluded largely for Type-II 2HDM, except for the tiny allowed region around $\tanb\in(8,10)$ under the combined direct and indirect searches of current LHC data at tree level. The excluded region is also nearly independent of parameter $m_{12}$ or $\lambda v^2=m_A^2-m_{12}^2/(\sin \beta \cos \beta)$. For larger $m_A$, the constraints get weaker, and direct searches can not put any more constraints on the wrong-sign region for $m_A=1500 \gev$.

The excluded region would change much after including loop corrections to the indirect searches. From \autoref{fig:tbcba_loopana}, the $sign(\kappa_b)=-1$ region could be corrected magnificently in some parameter space. Unlike the results in tree level, $m_{12}$ or $\lambda v^2$ could also make a difference. From \autoref{fig:tbcba_loop_sum}, we can conclude that the wrong-sign region with $\lambda v^2 <0$ will be stronger constrained and $\lambda v^2 >0$ will be less constrained under the combined constraints for $m_A=800 \gev$. 
For $m_A=1500 \gev$, region of $\lambda v^2\leq 0$ is totally excluded, and for large $\lambda v^2 >50 \gev^2$ the allowed region is shifted to right of the tree-level region to be less constrained. In general, we can conclude that with loop corrections, small $\lambda v^2$ will be more constrained  compared to tree level results,  while large $\lambda v^2$ is less constrained with current LHC limits. These features are quite different to the results at tree level. In \autoref{fig:mphim12_loop}, we explored the relationship $\cosba=2/\tanb$, and there are still allowed regions under current LHC and HL-LHC precision measurements, but when considering the future CEPC, it is difficult to find out the survived points.
%

  
\acknowledgments
We thank Huayang Song for useful discussions. This work is supported  by  the  Australian  Research  Council  Discovery  Project DP180102209.


\appendix



\bibliographystyle{JHEP}
\bibliography{ref_wrong-sign}

\end{document}